\documentstyle[12pt,epsfig]{kluwer}

\begin{document}
\begin{article}
\begin{opening}

\title{Parameters of the Magnetic Flux inside Coronal Holes}

\author{Valentyna \surname{Abramenko} and Vasyl \surname{Yurchyshyn}}
\institute{ Big Bear Solar Observatory, Big Bear City, CA 92314, USA}

\author{Hiroko \surname{Watanabe}}
\institute{Kwasan and Hida Observatories, Kyoto University, Kitakazan,
Ohminecyou, Yamashina, Kyoto 607-8471, Japan}

\begin{abstract}
Parameters of magnetic flux distribution inside low-latitude coronal holes (CHs)
were analyzed. A statistical study of 44 CHs based on Solar and Heliospheric
Observatory (SOHO)/MDI full disk magnetograms and SOHO/EIT 284\AA~ images showed
that the density of the net magnetic flux, $B_{{\rm net}}$,  does not correlate
with the associated solar wind speeds, $V_x$. Both the area and net flux of CHs
correlate with the solar wind speed and the corresponding spatial Pearson
correlation coefficients are 0.75 and 0.71, respectively. A possible explanation
for the low correlation between $B_{{\rm net}}$ and $V_x$ is proposed. The
observed non-correlation might be rooted in the structural complexity of the
magnetic field. As a measure of complexity of the magnetic field, the filling
factor, $ f(r)$, was calculated as a function of spatial scales. In CHs,  $f(r)$
was found to be nearly constant at scales above 2 Mm, which indicates a
monofractal structural organization and smooth temporal evolution. The magnitude
of the filling factor is 0.04 from the Hinode SOT/SP data and 0.07 from the
MDI/HR data. The Hinode data show that at scales smaller than 2 Mm, the filling
factor decreases rapidly, which means a mutlifractal structure and highly
intermittent, burst-like energy release regime. The absence of necessary
complexity in CH magnetic fields at scales above 2 Mm seems to be the most
plausible reason why the net magnetic flux density does not seem to be related
to the solar wind speed: the energy release dynamics, needed for solar wind
acceleration, appears to occur at small scales below 1 Mm.

\end{abstract}

\keywords{Sun: magnetic fields, coronal holes, solar wind}
\end{opening}

\section{Introduction}

Since the Skylab mission in early 1970s, it is believed that the coronal holes
seen on the surface of the Sun are related to the high-speed streams of the
solar wind and they might be a cause of geomagnetic disturbances (see, {\it
e.g.}, Sheeley {\it et al.} (1976) and references therein). A possibility of
ground-based observations of coronal holes in the spectral line He 1083.0 nm
(Harvey {\it et al.}, 1975) stimulated the interest to the problem. Numerous
sophisticated models were proposed to explain the CHs formation and evolution
({\it e.g.}, Wang and Sheeley, 1991; Fisk, 1996, 2001, 2005; Fisk {\it et al.},
1999; Wang {\it et al.}, 2000; Schrijver, 2001; Schrijver and Title, 2001;
Schrijver {\it et al.}, 2002; Schrijver and DeRosa, 2003).

During the last decade, a connection between the coronal holes and various
heliospheric phenomena, such as, high speed streams, corotating interaction
regions, long-living geomagnetic storms without CMEs (see, {\it e.g.}, Vrsnak
{\it et al.}, 2007 for references), became well established and stimulated
elaboration of approaches to forecast the geomagnetic response to the transit of
a coronal hole over the solar disk. An area occupied by coronal holes turned to
be a very fruitful parameter. Robbins {\it et al.} (2006) suggested an empirical
model to predict the solar wind speed at 1 AU from measurements of the
fractional area occupied by a CH inside a 14$^\circ$ - sectoral region centered
at the central meridian. A similar technique was independently applied later by
Vrsnak {\it et al.} (2007) and further extended to forecast, along with the solar
wind speed, other parameters of the solar wind, such as proton density,
temperature and magnetic field strength at 1 AU.

Several groups analyzed properties of the magnetic fields inside coronal holes
({\it e.g.}, Harvey and Sheeley, 1979; Harvey {\it et al.}, 1982; Obridko and
Shelting, 1989; Bumba {\it et al.}, 1995; Obridko {\it et al.}, 2000; Wiegelmann
and Solanki, 2004; Wiegelmann {\it et al.}, 2005; Abramenko {\it et al.}, 2006;
Hagenaar {\it et al.}, 2008). The most comprehensive study made so far on the
net magnetic fluxes and averaged flux densities in coronal holes was presented
by Harvey and colleagues (Harvey {\it et al.}, 1982), which was based on the
Kitt Peak full disk magnetograms (1 arcsec pixel size) and hand-drawn maps of
coronal holes derived from the He 1083.0 nm data. These authors analyzed the
ascending phase of the 21st cycle and reported an increase in the averaged net
flux density as the solar activity intensified. This was explained by an extra
flux deposited into low-latitude coronal holes by decaying active regions. Since
then, the density of the net flux inside CHs was adopted as a representative
characteristic of the magnetic filed inside coronal holes.

As long as it is the magnetic field that is ultimately responsible for
energetics in a coronal hole, it would be interesting to explore how the density
of the net magnetic flux is related to the speed of the fast solar wind. We
performed such a statistical study on the basis of the MDI full disk
magnetograms (Section 2) and found a rather surprising result: the density of
the net flux is not correlated with the solar wind speed measured at 1 AU. What
could be a reason for that? We suggested that the reason might be that the
density of the magnetic flux, derived from low-resolution data and averaged over
the CH's area, is not a suitable parameter to qualify energetics in CHs. The
ultimate reason for the observed non-correlation might stem from the structural
organization of the magnetic flux and its multifractal nature at small scales.
This encouraged us to study multifractal properties of the magnetic flux inside
a coronal hole taking advantage of the high resolution Hinode observations of
the magnetic field (Section 3). Our final section represents summary and
discussion of the results.

\section{ Averaged Net Flux Density Versus the Solar Wind Speed}

\subsection{Event Selection }

Here we focus on the distribution of the magnetic flux density of 44 CHs
observed between March 2001 and July 2006 at low solar latitudes near the center
of the solar disk (Table I). To avoid influence of the projection effect, we
required that the angular distance, $\theta$, from the disk center to the center
of gravity of a CH should not exceed 20 degrees. The values of $\theta$
(positive when the gravity center was located in the northern hemisphere) are
shown in the 8-th column of Table I. In the process of event selection we
discarded all CHs that had more than 5\% of their area outside a circle of 30
degree radius centered at the solar disk center.

We also made sure that the solar wind speed profiles were not contaminated by a
possible influence of ICMEs. Following Arge {\it et al.} (2004), we selected
only the events when, at the arrival time of the streams, the value of the
plasma beta \\
($http://omniweb.gsfc.nasa.gov/form/dx1.html$) was well above 0.1.

For the selected 44 CHs (Table I), we utilized the following data sets: i)
Michelson Doppler Imager (SOHO/MDI; Scherrer {\it et al.}, 1995) 1 min averaged
full disk magnetograms (spatial resolution 4 arcsec and the pixel size of 2
arcsec); ii) Fe {\sc xv} 284 \AA~ images from the EUV Imaging Telescope
(SOHO/EIT; Delaboudiniere {\it et al.}, 1995), and iii) ACE/SWEPAM and MAG
measurements of the solar wind at 1 AU. 

Each CH was determined in an EIT Fe {\sc xv} 284 \AA~ image as an area with
pixel
intensity below a certain threshold, namely, 80 DN. The threshold level was
first determined by trial-and-error method for one CH and it remained the same
for all the CHs analyzed here. Next, we co-aligned the EIT image, taken at the
time when the CH passed the central meridian with the closest MDI magnetogram.
The CH boundary was then mapped on that magnetogram, and calculations of the
magnetic field statistics were performed for an ensemble of pixels enclosed by
this boundary (Figure 1). In cases when the time difference between the EIT
image and the corresponding magnetogram exceeded 1/2 hour, we aligned the images
taking into account the differential rotation of the Sun. We discarded all
events where the time difference between the EIT and MDI images exceeded 4
hours. 

One more criteria was used to ensure that a given CH is indeed associated with
the feature observed in the solar wind speed profile. In case of the positive
association, the polarity of the $B_x$ component of the solar wind magnetic
field, measured in the GSE coordinate system by ACE/MAG, should be opposite to
that of the open flux in the base of the CH because the $x$-axis in the GSE
system points from the Earth toward the Sun, while the positive magnetic field
on the solar surface is directed outward from the Sun. Only those CHs, for which
magnetic polarities satisfied the above condition, were included into this
study.

 \begin{figure}[!h]\centerline{\epsfig{file=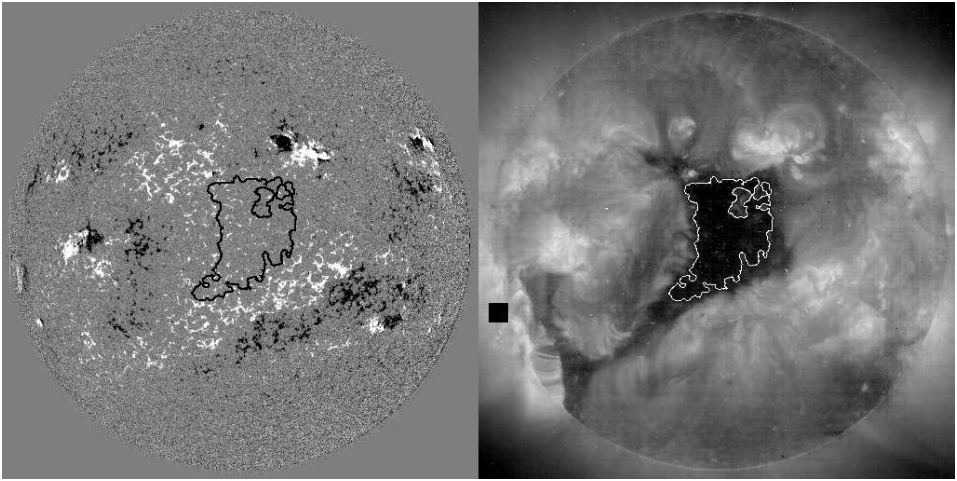,width=30pc}}
\caption{SOHO/MDI magnetogram (left) and the corresponding SOHO/EIT 284 \AA
 image taken at 19:06 UT on 2 March 2002 (right). The contour indicates
the boundary of coronal hole CH201 (see Table I).}
\label{fig1}
\end{figure} 

\subsection{Solar Wind Speed Derivation}

For each CH in our data set, we determined the solar wind speed, $V_x$. We
utilized the 64 s averaged time profiles measured in the GSE coordinate system
with ACE/SWEPAM instruments. Note that the solar wind speed acquires negative
values in the GSE coordinate system. 

The arrival time at 1 AU of the solar wind associated with the CH was
determined as the moment $t$, when a function
\begin{equation}
\Delta r(t) = V_x (t-t_0) - 1 {\rm AU}
\label{Dr}
\end{equation}
turns into zero. Here, $t_0$ (3rd column of Table I) is the CH culmination time,
and $t=t_{\rm A}$ (4-th column of Table I) is the arrival time.

\begin{figure}[!h]\centerline{\epsfig{file=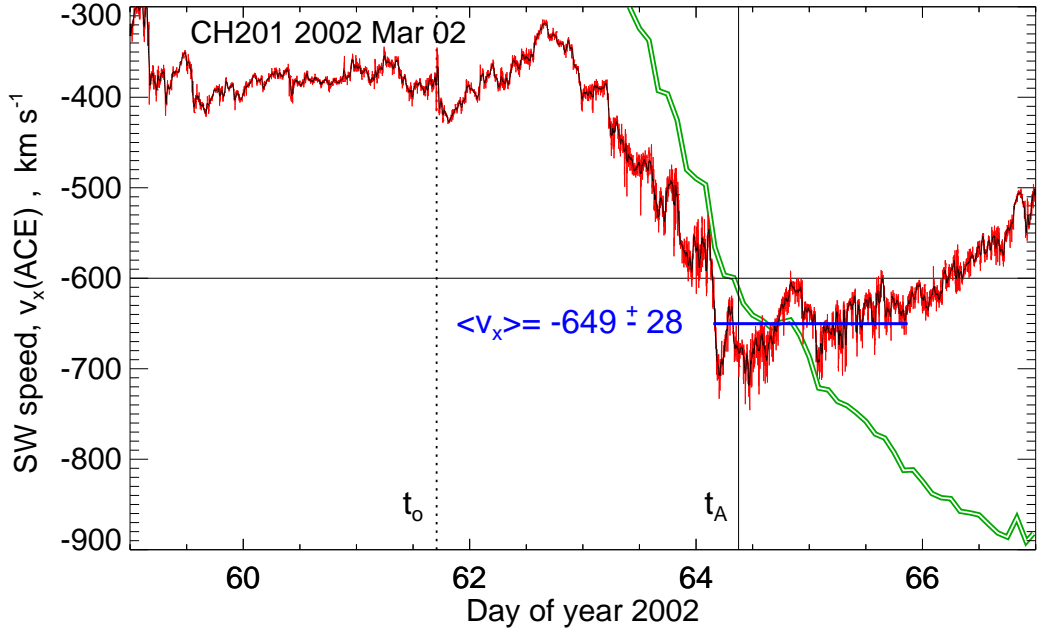,width=38pc}}
\caption{Observed time profile of the $V_x$ component of the solar wind speed
(red curve) as measured by ACE satellite from 0000UT on 28 February 2002 until
2359UT on 7 March 2002. The black curve represents a smoothed over 21-point
observed speed profile. The vertical dotted line indicates the CH culmination
time, $t_0$, while the vertical solid line indicates the solar wind arrival
time, $t_{\rm A}$. The arrival time was determined as the moment when parameter
$\Delta r$ (double green curve) is equal to zero. The blue thick horizontal line
segment shows the time interval used to determine the average, for a given
event, solar wind speed, $\langle V_x \rangle $ (indicated by the vertical
position of the blue line segment).}
\label{fig2}
\end{figure}

We then accepted that the association between the CH and the solar wind feature
is reliable when $t_{\rm A}$, calculated from Equation (\ref{Dr}), falls into
the well pronounced ``dip'' in the observed solar wind speed profile. Figure
\ref{fig2} shows an observed time profile of the solar wind speed and the
corresponding profile of $\Delta r(t)$ associated with the CH shown in Figure
\ref{fig1}.

The magnitude of the solar wind speed, $\langle V_x \rangle$, was determined by
averaging the observed time profile of $V_x(t)$ over a time interval that
includes the minimum of the observed speed profile. This time interval (marked
by the thick blue horizontal segment in Figure \ref{fig2}) was defined in the
following way. First, we applied a 21-point running averaging to the observed
$V_x(t)$ profile (red curve in Figure \ref{fig2}). This smoothed profile (black
curve) was then used to determine i) the level of undisturbed slow-wind speed,
$V_0$, preceding the CH, and ii) the maximum speed inside the CH, $V_{{\rm
max}}$. To obtain the averaged solar wind speed, $\langle V_x \rangle$, we
averaged all data points inside an interval where $V_x(t)$ is lower than
$V_{{\rm max}} + (\vert V_{{\rm max}} \vert - \vert V_0 \vert)/4$. This
threshold was chosen by a trial-and-error method. The resulting averaged solar
wind speed is indicated in Figure \ref{fig2} by the vertical position of the
blue line segment. We would like to emphasize here that we did not include in
our study those CHs whose associated speed profile was so complex that no well
defined minimum could be chosen.

For each CH in our data set, the area of the CH (in pixels of the MDI full disk
magnetogram) and the associated solar wind speed are listed in the 7-th and 9-th
columns of Table I, respectively. 

The above routine allowed us to determine the arrival time and the magnitude of
the solar wind speed individually for each coronal hole. This technique differs
from that applied by Robbins {\it et al.} (2006) and Vrsnak {\it et al.} (2007),
where the mean arrival time, determined from the cross-correlation technique,
was utilized.

\subsection{Magnetic Parameters }

The above selection routine left us with a set of 44 CHs. MDI full disk 1 minute
average magnetograms were available for all of them. However, the full disk data
of 1 minute cadence, as well as the observations in the MDI high resolution
mode, were not available for all of them. Since we are interested in
measurements of the net magnetic flux density, we decided first to focus on
whether the measurements of this parameter from 1 minute full disk magnetograms
are reliable. To do that, we analyzed coronal hole CH 204 from our data list
(see Table I), for which nearly simultaneous observations in two MDI modes were
available: 1 minute cadence full disk and high resolution modes. Inside this
large coronal hole, located at the solar disk center, we outlined a rectangular
area of $n \times m$ pixels and calculated magnetic parameters for five
different magnetograms. Results are compiled in Table II, where the first row is
the result from one original individual MDI full disk 1 minute magnetogram. The
second row represents results from the same magnetogram, but smoothed in both
directions with a boxcar average of three pixels. The third row are parameters
from a magnetogram derived from averaging of five consecutive original 1 min
magnetograms, whereas its smoothed version is presented in the next row. The
last row represents results from the high resolution MDI magnetogram recorded at
the same place on the Sun with pixel size of 0.6 arcsec. 

From each of the above magnetograms, we first calculated the net flux density,
$B_{{\rm net}}=\sum B_{||}/(n \times m)$ (2nd column in Table II), inside the
box. We note that the net flux is obtained by summing all pixel values with
their sign. This procedure results in cancellation of the bulk of sign-variable
noise. Magnetic flux confined in loops, closed within the CH, will also be
canceled. This allowed us to assume that in CHs, the net flux density also
represents the open magnetic flux density.

The density of the total unsigned flux, $B_{{\rm tot}}=\sum |B_{||}|/(n \times
m)$, is shown in the 3rd column of Table II. The 4-th column represents the
density, $B_{\rm n}$, of the noise in magnetograms derived from the power spectrum
calculations as it was suggested by Longcope and Parnell (2008). To derive this
parameter, we calculated a 1D power spectrum, $S(k)$, from each magnetogram and
then calculated the noise density as $B_{\rm n}=(\pi k_c^2S(k_c))^{1/2}$, where
$k_c=2\pi^{1/2}\Delta x$ is the maximum wave number. The imbalance of the
magnetic flux is shown in the 5-th column. The 6-th column represents the
density of the net flux calculated over the area where $|B_{||}|<3 B_{\rm n}$ (we
denote $3 B_{\rm n}=p$), and the last column shows the fraction of the box area,
where the flux density exceeds the triple noise level, $|B_{||}|>p$.

Data of Table II indicate that the density of the net flux, $B_{{\rm net}}$,
does not vary much (by less than 2\%) with temporal/spatial averaging and does
not depend on the noise level. At the same time, the density of the unsigned
flux, $B_{{\rm tot}}$, depends significantly on the noise level. (Note that the
magnitude of $B_{\rm n}$ is in good agreement with the results from MDI full
disk data noise reduction reported by Hagenaar {\it et al.} (2008)). Last two
columns in Table II show that the bulk of the magnetic flux in CHs is
concentrated within a small (a few percents) fraction of the CH's area. Vast
zones of low fluctuations contribute only about 1 G into the resulting magnitude
(4 G) of the net flux density.

This experiment shows, first, that an averaging procedure is undesirable when
one intends to analyze fine structures of the magnetic field. Second, that the
net flux density in coronal holes is measured with the same level of confidence
from the original and averaged magnetograms. High noise level in original
magnetograms does not influence much on the $B_{{\rm net}}$ calculations. The
reasons for that are: i) high imbalance of the magnetic flux inside coronal
holes, accompanied by low (as compared to adjacent quiet sun areas) rate of
dipole emergence (Abramenko {\it et al.}, 2006; Hagenaar {\it et al.}, 2008); 
and ii) magnetic features that contribute to the open flux are predominantly
well above the noise level, even for the noisiest magnetogram. 

For 44 coronal holes analyzed here, magnitudes of the net magnetic flux,
$\Phi_{{\rm net}}= \sum B_{||} \Delta s$, where $\Delta s$ is a pixel size, are
presented in the 5-th column of Table I. The 6-th column shows the densities of
the net magnetic flux, $B_{{\rm net}}=\sum B_{||}/A$, where $A$ is the area in
pixels inside the CH boundary, 7-th column in Table I. We would like to
emphasize that the moduli of $B_{{\rm net}}$ tend to decrease from March 2001
toward July 2006, {\it i.e.}, toward the end of the 23rd cycle. This is in
qualitative agreement with Harvey {\it et al.} (1982).

\subsection{ Correlations Between Calculated Parameters}

\begin{figure}[!h]\centerline{\epsfig{file=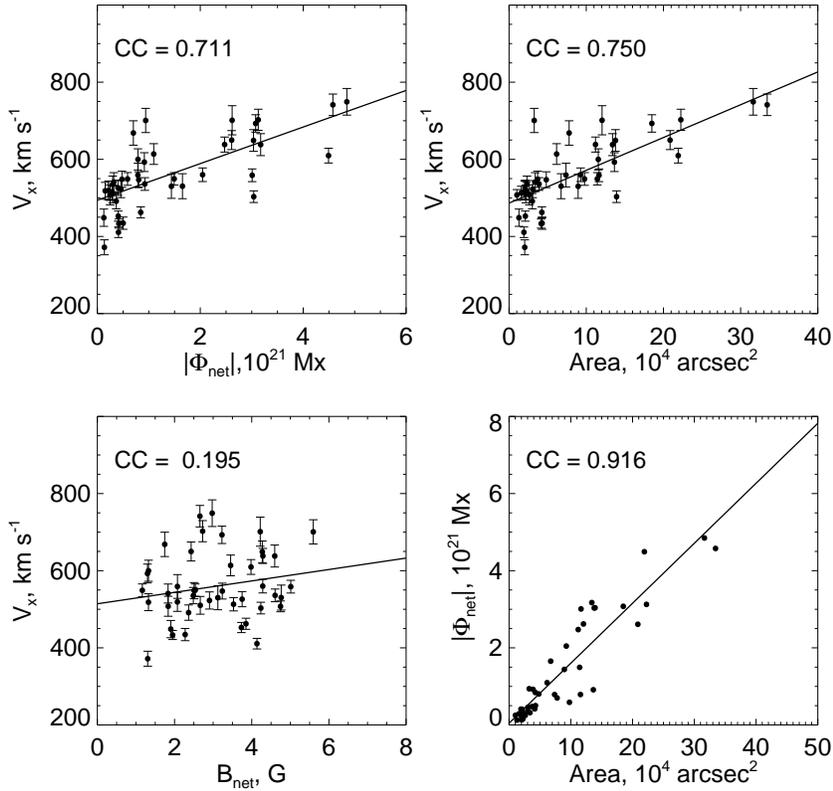,width=27pc}}
\caption{Correlations between magnetic parameters of coronal holes and solar
wind parameters. Solid straight lines represent the best linear fit to the
data points.}
\label{fig3}
\end{figure}

Figure \ref{fig3} shows correlations between parameters of CHs and the solar
wind speed. The solar wind speed, $\langle V_x \rangle$, is directly
correlated with the total net flux (upper left panel). The corresponding Pearson
coefficient is 0.71 and the linear fit is given by
\begin{equation}
\langle V_x \rangle = 494 + 47.4 \cdot \Phi_{{\rm net}},
\end{equation}
where $\langle V_x \rangle$ is in km s$^{-1}$ and $\Phi_{{\rm net}}$ is in 
$10^{21}$ Mx.

However, the total net flux is a product of the CH area, $A$, and the average
net flux density, $B_{{\rm net}}$. According to Figure \ref{fig3}, the solar
wind speed is strongly correlated with the CH area (upper right panel,
correlation coefficient (cc)=0.75) and only weekly depends on the averaged net
flux density (lower left panel, cc=0.20). Linear fitting to the data point gave
us the following relationship between the speeds and CH areas:
\begin{equation}
\langle V_x \rangle = 486 + 8.50 \cdot A,
\end{equation}
where $A$ is in $10^4$ arcsec$^2$. As a result, there is a very strong
dependence of the total net magnetic flux from the CH area (cc=0.92, lower right
panel), which allows a reliable estimation of the total open flux in a CH:
\begin{equation}
\Phi_{{\rm net}} = 0.045 + 0.156 \cdot A,
\label{fop}
\end{equation}
where $\Phi_{{\rm net}}$ is in $10^{21}$ Mx  and $A$ is in $10^4$ arcsec$^2$.

These findings very well agree with previously reported results (Wang and
Sheeley, 1990; Robbins {\it et al.}, 2006; Vrsnak {\it et al.}, 2007; Schwadron
and McComas, 2008) in spite of differences in used techniques. This consistency
indicates reliability of the coronal hole analysis presented here.

On many occasions, analyzed CHs had closed loop structures embedded within
them, which, in general, reduces the area occupied by the open flux and may lead
to different magnitudes of the averaged total flux density. We excluded the
closed loop areas from our calculations and repeated the above analysis. As
expected, since the closed loop regions substitute a small fraction of the CH
area, accounting for them did not lead to a significant change of the
relationship between the solar wind speed and the total flux density. In fact,
the Pearson coefficient even slightly decreased to 0.15.

We finally would like to note that the well defined relationship between
the speed and area opens up a possibility to estimate the solar wind speed 3-4
days in advance by measuring the area of a CH, when it passes the central
meridian. (Note that in this study the CH area was measured inside a contour of
80 DN in EIT/Fe {\sc xv} 284 \AA~ images.)

\begin{table}[!ht]
\caption{\sf List of studied coronal holes and the corresponding parameters.}
\footnotesize
\begin{center}
\begin{tabular}{lrrrrrrrrr}
\hline
CH   & Culmination &$t_0$,&$t_{\rm A}$,&   $\Phi_{{\rm net}}$,& $B_{{\rm net}}$
&
 $A$,& $\theta$,& $\langle V_x \rangle$,   \\ name & time  &  doy & doy  
   &  $10^{21}$ Mx         & G &  pixels &  deg     & km s$^{-1}$      \\

\hline
CH191    &2001Mar02/23:00&      61.96&  64.96&   3.00   &  5.01	& 29145&  17.2&   559 $\pm$16 \\
CH230	 &2002Feb03/22:00&	34.92&	37.75&   1.09   &  3.43	& 15408&  -2.8&   614 $\pm$27 \\
CH201	 &2002Mar02/17:00&	61.71&	64.38&   3.03   &  4.27	& 34472&  2.2&	 649 $\pm$28 \\
CH202	 &2002Mar29/18:00&	88.75&	91.08&   2.61   &  4.22	& 30127&  1.7&	 701 $\pm$38 \\
CH210	 &2002Apr28/23:00&	118.96&	122.63&  0.84   &  3.85	& 10621&  -19.6&  462 $\pm$14 \\
CH231	 &2002Jul03/12:00&	184.5&	187.67& -0.92   & -4.60 &  9721&  -13.2&  536 $\pm$16 \\
CH204	 &2002Nov01/17:00&	305.71&	309.04&  1.49   &  2.54	& 28578&  -6.3&   549 $\pm$16 \\
CH350	 &2003Jan27/06:00&	27.25&	30.67&  -0.24   & -4.75	&  2555&   1.6&	 507 $\pm$14 \\
CH302	 &2003Feb23/19:00&	54.79&	58.13&  -2.04   & -4.26 & 23176&  7.1&	 560 $\pm$18 \\
CH303	 &2003Mar01/03:00&	60.13&	63.29&  -1.65   & -4.76	& 16856&  14.2&   530 $\pm$32 \\
CH304	 &2003Apr08/09:00&	98.38&	100.79&  3.12   &  2.73	& 55633&  -7.9&   702 $\pm$27 \\
CH305	 &2003Apr24/07:00&	114.29&	117.71& -0.29   & -2.67	& 5407&   0.2&	 510 $\pm$23 \\
CH307	 &2003May03/19:00&	123.79&	126.38&  3.07   &  3.23	& 46245&  -11.1&  693 $\pm$23 \\
CH354	 &2003May21/01:00&	141.04&	144.54& -1.43   & -3.12	& 22381&  -1.0&   530 $\pm$31 \\
CH309	 &2003May25/19:00&	145.79&	148.21& -0.93   & -5.59	& 8151&   -20&	 701 $\pm$31 \\ 
CH312	 &2003Jun11/11:00&	162.46&	165.63& -0.47   & -2.51	& 9260&   -0.6&   548 $\pm$21 \\
CH355	 &2003Aug10/01:00&	222.04&	224.63& -2.61   & -2.43	& 52146&  3.6&	 650 $\pm$25 \\
CH314	 &2003Aug20/19:00&	232.79&	235.04&  4.57   &  2.66	& 83592&  -4.9&   741 $\pm$28 \\
CH356	 &2003Aug30/03:00&	242.13&	245.29& -0.40   & -3.75	& 5191&   7.8&	 526 $\pm$20 \\
CH357	 &2003Sep07/23:00&	250.96&	253.63& -3.17   & -4.60	& 33515&  -19.2&  638 $\pm$28 \\
CH315	 &2003Sep16/08:00&	259.33&	261.5&	 4.84   &  2.98	& 79079&  -4.2&   749 $\pm$35 \\
CH317	 &2003Dec01/05:00&	335.21&	339.21&  0.12   &  1.90	& 3207&   -6.0&   449 $\pm$22 \\
CH320	 &2003Dec18/19:00&	352.79&	355.79& -4.49   & -3.98 & 54777&  6.3&	 610 $\pm$19 \\
CH455	 &2004May17/10:00&	138.42&	141.75&  0.21   &  2.07	& 5074&   -2.7&   519 $\pm$24 \\
CH456	 &2004May28/23:00&	149.96&	153.13& -0.58   & -1.17	& 24452&  13.0&   549 $\pm$17 \\
CH457	 &2004May31/03:00&	152.13&	155.54& -3.03   & -4.24	& 34826&  -3.3&   503 $\pm$15 \\
CH458	 &2004Jun05/01:00&	157.04&	160.96& -0.40   & -3.73	& 5281&   14.2&   452 $\pm$13 \\
CH459	 &2004Jun12/10:00&	164.42&	167.75&  0.31   &  1.83	& 8469&   -9.8&   542 $\pm$25 \\
CH460	 &2004Jul07/23:00&	189.96&	193.71&  0.36   &  2.36	& 7540&   11.4&   491 $\pm$20 \\
CH461	 &2004Jul13/13:00&	195.54&	199.13&  0.15   &  1.32	& 5590&   2.8&	 518 $\pm$22 \\
CH462	 &2004Nov21/19:00&	326.79&	330.21&  0.28   &  3.53	& 3967&   16.0&   513 $\pm$17 \\
CH412	 &2004Nov27/17:00&	332.71&	335.63& -2.47   & -4.29	& 27994&  -1.2&   638 $\pm$19 \\
CH502	 &2005Feb15/07:00&	46.29&	49.54&  -0.80   & -3.24	& 12051&  14.8&   547 $\pm$21 \\
CH503	 &2005Apr17/02:00&	107.08&	110.42&  0.45   &  2.90	& 7650&   -9.5&   523 $\pm$22 \\
CH554	 &2005May29/05:00&	149.21&	153.63&  0.50   &  2.28	& 10710&  -14.0&  434 $\pm$16 \\
CH558	 &2005Jul26/10:00&	207.42&	210.58& -0.78   & -2.08	& 18438&  -5.4&   559 $\pm$31 \\
CH563	 &2005Oct11/19:00&	284.79&	289.63&  0.13   &  1.30	& 5075&   1.9&	 372 $\pm$19 \\
CH564	 &2005Oct23/15:00&	296.63&	299.88&  0.24   &  1.83	& 6557&   -1.9&   508 $\pm$26 \\
CH567	 &2005Dec26/15:00&	360.63&	363.04& -0.70   & -1.75	& 19465&  2.3&	 668 $\pm$32 \\
CH601	 &2006Jan12/19:00&	12.79&	16.71&   0.41   &  1.95	& 10396&  4.3&	 433 $\pm$12 \\
CH603	 &2006Feb12/03:00&	43.13&	46.79&   0.29   &  2.48	& 5799&   -8.3&   535 $\pm$21 \\
CH604	 &2006Feb24/19:00&	55.79&	60.21&   0.40   &  4.14	& 4813&   -4.9&   411 $\pm$14 \\
CH606	 &2006May03/09:00&	123.38&	126.88&  0.78   &  1.33	& 28904&  -4.0&   600 $\pm$27 \\
CH607	 &2006Jul02/19:00&	183.83&	186.67& -0.91   & -1.30	& 34120&  2.0&	 593 $\pm$24 \\
\hline
\end{tabular}
\end{center}

\end{table}


\begin{table}[!ht]
\caption{\sf SOHO/MDI magnetic measurements for a box inside CH204 coronal
hole.}
\footnotesize
\begin{center}
\begin{tabular}{lcccccc}
\hline
Type of & $B_{{\rm net}}$ & $B_{{\rm tot}}$ & $B_{\rm n}$& Flux
&$B_{{\rm net}}(<p)$& Area above\\ 
 magnetogram & G & G & G & imbalance, \% & G & noise level, \% \\
\hline
Full disk 1 min & 4.53 & 19.5  & 17.3  & 23 &1.11  &4.1  \\
\hline
Full disk 1 min &&&&&&\\
smoothed        & 4.50  & 10.3  & 4.17  & 44  &0.38  &19  \\
\hline

Full disk 5 min ave & 4.48 & 12.1 & 9.3 & 37  &0.84 &6.1  \\
\hline
Full disk 5 min ave &&&&&& \\
smoothed            & 4.48 & 8.1 & 3.2 & 55 &0.23 &16 \\

\hline
Hi res          & 3.88 & 12.4 & 8.9 & 31 &1.64 &6.6  \\
\hline

\end{tabular}
\end{center}

\end{table}

\section { Multifractality in the Magnetic Field in Coronal Holes}

The absence of a relationship between the solar wind speed and the density of
the magnetic flux, seems to be suspicious at the first sight. The solar wind
speed represents the intensity of the solar wind acceleration processes, {\it
i.e.}, energy release dynamics, which ultimately is related to the magnetic
field. Numerous recent studies of the dynamics of energy release in the low
corona based on Hinode data (Baker {\it et al.}, 2008; Suematsu, 2008; Shimojo,
2008; Moreno-Insertis {\it et al.}, 2008) tell us that there
is a role for the magnetic field to play in these phenomena. Two things could be
done to resolve the problem. First, improve spatial resolution of magnetic field
measurements, and second, explore the proper measures of complexity of the
magnetic field.

\begin{figure}[!h]\centerline{\epsfig{file=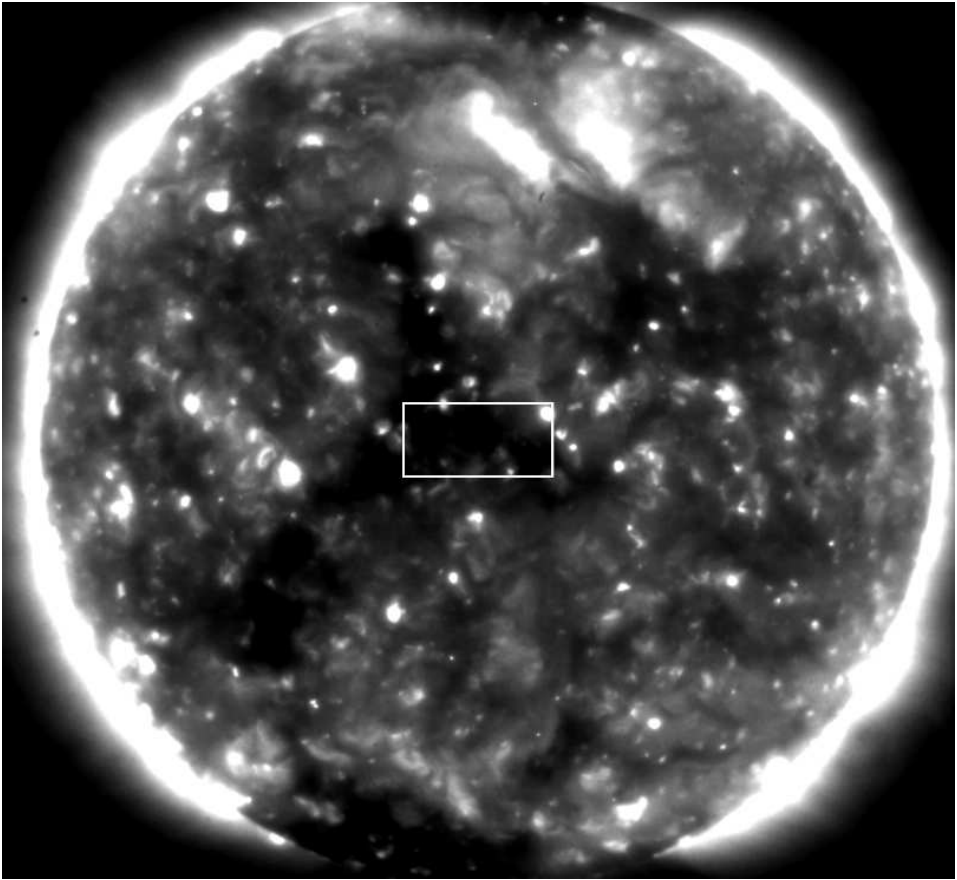,width=30pc}}
\caption{ The 11:42UT XRT image on 30 November 2008.  The box in the
center of the solar disk outlines the area inside a coronal hole where the
 SOT/SP magnetogram was taken (Figure \ref{fig5}). } 
\label{fig4}
\end{figure}

\begin{figure}[!h]\centerline{\epsfig{file=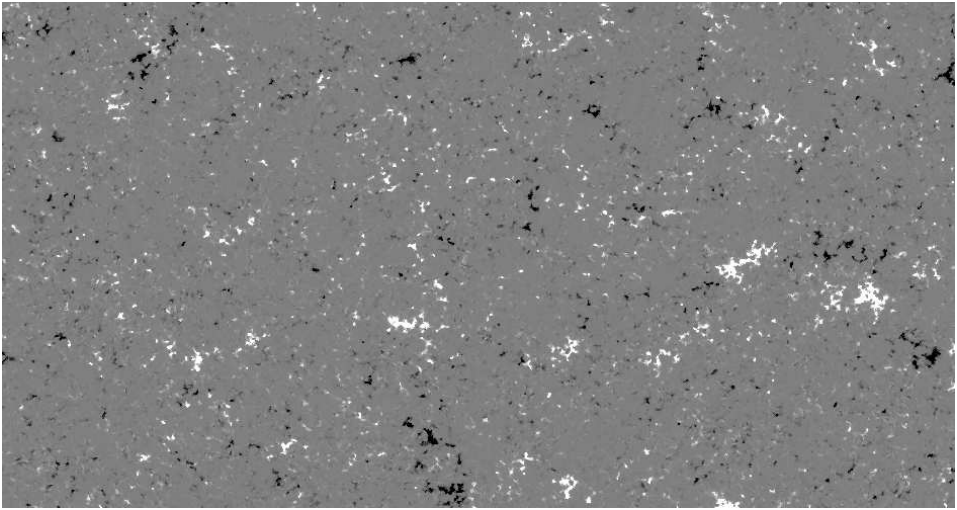,width=30pc}} \caption{Hinode
SOT/SP magnetogram of the coronal hole area outlined in Figure \ref{fig4}
obtained in the fast mode (pixel size of 0.297$\times$0.320 arcsec) and
calibrated with the HAO inversion code. The magnetogram was despiked. The size
of the solar surface is 200$\times$115 Mm. The image is scaled in the range from
-500 to 500 G.} 
\label{fig5}
\end{figure}

The Solar Optical Telescope (SOT, Tsuneta {\it et al.}, 2008) onboard {\it
Hinode} has a 50 cm aperture mirror and is the largest optical solar telescope
ever sent to space. The Hinode/Spectropolarimeter (SP, Ichimoto {\it et al.},
2008) obtains the full Stokes parameters using the Fe {\sc i}  630 nm absorption
line and offers a unique opportunity to obtain magnetograms with pixel size of
0.32 arcsec in the fast mode and 0.16 arcsec in the normal mode. Coronal holes
at low latitudes are rare during a solar minimum, where we are now.
Nevertheless, on 11 November 2008 the SOT/SP instrument recorded a CH (Figure
\ref{fig4}) at the disk center in the fast mode. We analyzed here one of the
SOT/SP magnetograms of 942$\times$500 pixels taken in the fast mode (Figure
\ref{fig5}). The inversion code and calibration routine were performed at HAO
CSAC ($http://www.csac.hao.ucar.edu/$). The magnetogram was carefully despiked.

It is believed now that magnetic structures in active regions and in quiet sun
areas are fractals (Tarbell {\it et al.}, 1990; Schrijver {\it et al.}, 1992;
Balke {\it et al.}, 1993; Lawrence {\it et al.}, 1993; Meunier, 1999; Ireland
{\it et al.}, 2004; Janssen {\it et al.}, 2003; McAteer {\it et al.}, 2005). In
particular, Meunier (1999) and McAteer {\it et al.} (2005) showed that magnetic
structures of active regions, even those recorded with 4 arcsec resolution (MDI
full disk magnetograms), are fractals. Janssen {\it et al.} (2003) showed that
the magnetic field in a quiet sun ares is also a fractal. Fractals are
self-similar, porous structures with jagged boundaries. Their scaling parameters
do not vary with  spatial scale. For example, the filling factor, {\it i.e.}, 
the ratio of the area occupied by (above-noise) fields to the entire area, does
not vary when the resolution changes. An example: a comparison of the 3rd and
5-th rows in the last column in Table II tell us that the fraction of the area
occupied by strong fields is nearly the same, about 6\%, for the 4 arcsec and
1.2 arcsec resolutions. This might indicate that the coronal hole magnetic field
 structure at these scales can be considered as a single fractal, or, in other
words, a {\it monofractal}.

Temporal variations in monofractal structures are non-intermittent, {\it i.e.},
high fluctuations in energy release are very rare and they do not represent a
burst-like behavior. Therefore, strong energy release events are rare and they
do not define the energy balance dynamics.

Another situation is when a multifractal is formed. Multifractals form in nature
ubiquitously when several processes contemporaneously govern formation of a
structure, each one dictating its own rules of clustering and fragmentation. A
highly intermittent temporal behavior is inherent for multifractals, so that
high fluctuations in energy release are not rare, and the regime of violent,
burst-like energy release dynamics is set. Therefore, to relate the energy
release dynamics with the magnetic field, one should find the spatial scales
where multifractal properties of the magnetic field manifest themselves, if any.
(Note that the terms multifractality and intermittency describe the same
property of a structure. However, the former is related to the spatial domain
while the later is usually utilized for the temporal domain; see, {\it e.g.},
Abramenko (2008) for details.)

One of possible ways to diagnose multifractality is based on calculation of the
filling factor as a function of spatial scale (Frisch, 1995). For multifractals,
the filling factor decreases as the scale decreases. In other words, the
fraction of a volume occupied by strong fields (the so called active mode),
decreases as we study the multifractal at smaller and smaller scales. For
monofractals, this ratio is constant with scale. To this end, our goal was to
explore how the filling factor varies with scale inside coronal holes.

The ratio of the active mode to the entire volume as a function of the scale,
$r$, can be calculated from the flatness function, which for the
longitudinal magnetic field $B_{l}$, can be written as (Frisch, 1995): 
\begin{equation} 
F(r)=S_4(r)/(S_2(r))^2.
\label{Fr}
\end{equation} 
Here,
\begin{equation}
S_q(r) = \langle | {B_l}({\bf x} + {\bf r}) - {B_l}({\bf x})|^q \rangle 
\label{Sq} 
\end{equation}  
are the structure functions, and ${\bf x}$ is the current pixel on a
magnetogram, ${\bf r}$ is the separation vector ({\it i.e.}, the spatial scale),
and $q$ is the order of a statistical moment, which takes on real values. The
angular brackets denote averaging over the magnetogram.  Details of the
calculation routines and applications can be found in Abramenko {\it et al.} 
(2002, 2003, 2008) and in Abramenko (2005). The filling factor is then calculated as
\begin{equation}
f(r) = 1/F(r).
\label{fr}
\end{equation}
As it was mentioned above, the filling factor does not depend on the spatial
scale, $r$, in case of a  monofractal structure. On the contrary, for a 
multifractal, the filling factor displays a power-law decrease as the scales
become progressively smaller.

\begin{figure}[!h]\centerline{\epsfig{file=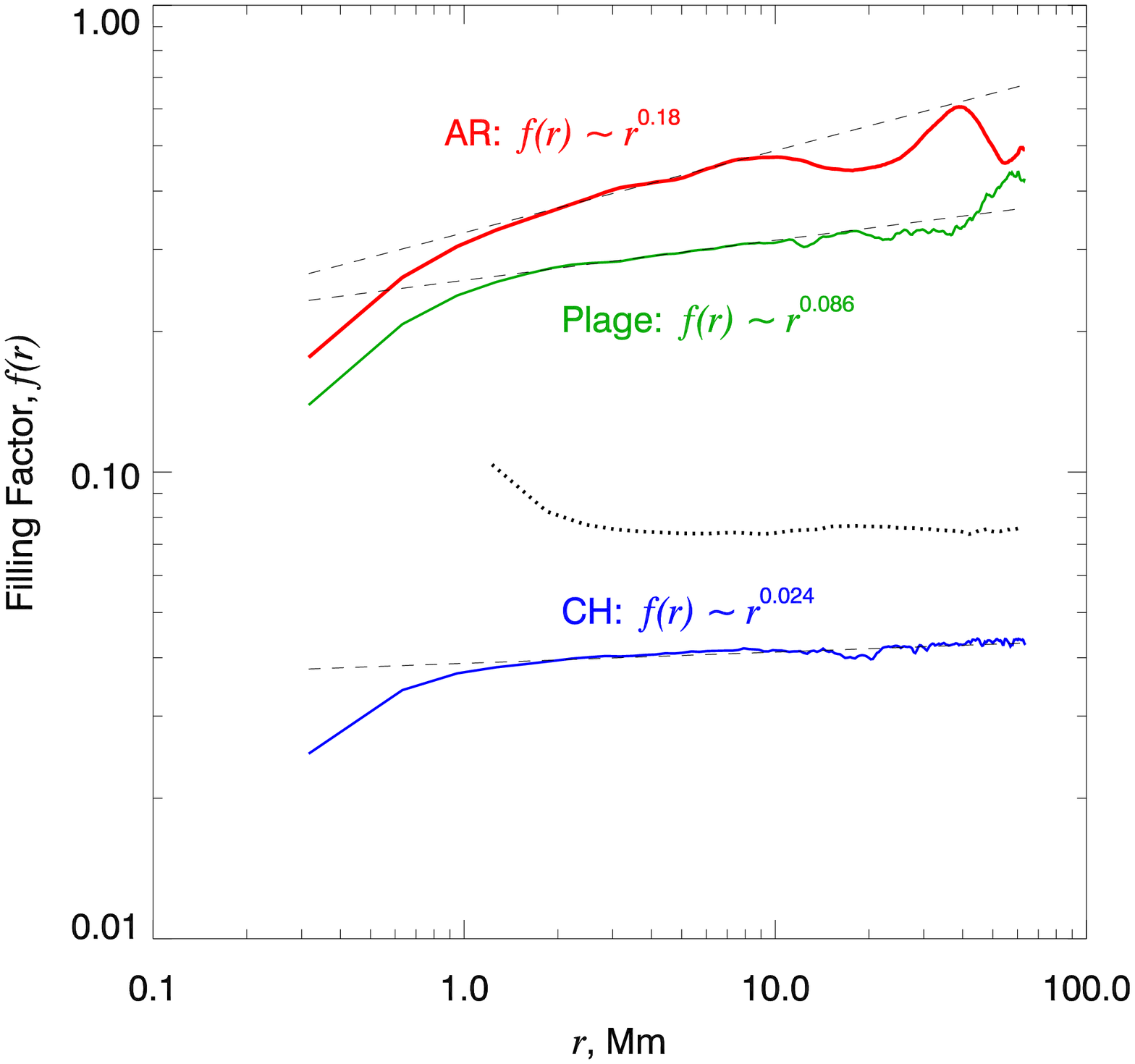,width=29pc}}
\caption{ Variation of the filling factor, $f$, with spatial scale, $r$. Blue -
data from the Hinode/SOT/SP magnetogram of the coronal hole shown in Figure
\ref{fig5}. Black dots - the averaged filling factor calculated from 36
magnetograms for 19 coronal holes observed between 2002 and 2004 with SOHO/MDI
in the high resolution mode. The flat filling factor is observed at scales above
3 Mm in both data sets. The decreasing of the filling factor at scales below 2
Mm is observed in SOT/SP data. For comparison, data for the active region NOAA
0930 (red) and  weak plage area (green) derived from the Hinode/SOT/SP fast mode
magnetograms are shown. The active region and the plage area display the
decreasing filling factor at large scales above 2 Mm. Dashed lines represent the
best linear fit to the data points inside an interval of $\Delta r=(1.6 - 8.2 )$
Mm. The steeper slope of the fit corresponds to higher degree of
multufractality.}
\label{fig6}
\end{figure} 

Results of the filling factor are presented in Figure \ref{fig6}.
Hinode/SOT/SP data for the CH observed on 30 November 2008 (blue curve)
indicates that at scales larger than approximately 2 Mm, the magnetic field
structure seems to be a monofractal. Only a very slight slope (0.024) of the
power law linear fit is observed. Thus, at scales larger than 2 Mm, the magnetic
field in a coronal hole seems to be a monofractal.

For comparison, we calculated and plotted in the same graph (Figure \ref{fig6})
the filling factor for an active region NOAA 0930 and a plage area to the west
of this active region. The corresponding magnetograms were also derived with the
SOT/SP instrument in the fast mode and processed with the same routines as those
applied for the CH magnetogram. Data for the AR and plage show a steeper slope
of the power law linear fit with indices of 0.18 and 0.09, respectively. The
dependence of the filling factor on the scale implies multifractality of the
magnetic field at scales larger than 2 Mm.

To double check our inference on monofractality in CHs at large scales, we
calculated the filling factor from the MDI/HR data for 36 square areas located
inside 19 coronal holes observed during 2002-2004 at the solar disk center. The
MDI results are shown in Figure \ref{fig6} with the dotted line. At scales
larger than approximately 3 Mm, the filling factor is constant, which confirms
our inference on the monofractality of the coronal hole magnetic field. At
spatial scales smaller than  $r \approx 3$Mm the filling factor function is
influenced by noise and resolution of the MDI data and it tends to grow as the
scale decreases. This effect is caused by the Gaussian nature of the data noise,
with its intrinsic filling factor of 1/3.

What is interesting is that the Hinode data do not show similar behavior of
the filling factor at the same scales, $r<3$ Mm. Instead, below $r=2$ Mm, the
Hinode SOT/SP filling factor displays a rapid decrease, which does not seem to
be caused by data noise. The decrease of $f(r)$ is well pronounced for the
coronal hole, an active region and plage area data. 

The break in the filling factor function is observed at scales approximately 
of 1 Mm. We may conclude that below this scale, the CH magnetic field is
a multifractal.

\section{Summary and Discussion }

In summary, analyzing magnetic fields inside the low-latitude coronal holes, we
arrived at the following conclusions.

i) The density of the net magnetic flux does not correlate with the
corresponding {\it in situ} solar wind speeds. At the same time, both CH area
and total net flux correlate very well with the solar wind speed and the
corresponding spatial Pearson correlation coefficients determined for 44 CHs are
0.75 and 0.71, respectively. The relationship between the CH areas and the solar
wind speed with almost the same results in correlation was earlier derived by
Robbins {\it et al.} (2006) and by Vrsnak {\it et al.} (2007).

As we discussed in Section 1, the fact that the net magnetic flux density 
is not correlated with the solar wind speed can hardly be interpreted as
irrelevance of the magnetic field to the solar wind acceleration process. One
would rather suggest that the net flux density measured with the resolution of
4 arcsec does not reflect the energy release dynamics inside
CHs. Analysis of SOHO/MDI/high-resolution and Hinode/SOT/SP magnetograms seems
to support this suggestion. 

ii) The filling factor as a measure of multifractality in CHs was calculated as
a function of spatial scale. It was found to be nearly constant at scales above
2 Mm. Its magnitude is approximately 0.04 from the Hinode data and 0.07 from the
MDI/HR data. At scales smaller than 2 Mm, the filling factor starts to decline
as the scale becomes smaller, and at approximately 1 Mm the regime of fast
decrease of the filling factor is set. 

A constant filling factor at $r>2$ Mm in the CH magnetic fields indicates 
their monofractal nature and self-similarity. A self-similar
structure by definition possesses constant statistical parameters at all scales
(such as various scaling exponents, including the filling factor). However, only
for artificial, mathematically created fractals, self-similarity is present at
infinite range of scales (see, {\it e.g.}, Schroeder, 1991). For a majority of
structures in nature, however, self-similarity with constant scaling parameters
only holds at a finite range of scales, while at the entire interval of scales
the scaling parameters are different. As a result, a multifractal structure
forms with a superposition of many fractals, each one imposing its own scaling
rules. A crucial difference between monofractals and multifractals is in their
temporal evolution. In monofractals, large fluctuations of parameters (say,
energy release events) are rare and do not determine mean values. In other
words, time profiles are non-intermittent and evolution proceeds without
catastrophes. On the contrary, in multifractals, the time profiles are highly
intermittent, large fluctuations are not rare, and they determine mean values.
The temporal energy release process is burst-like. 

If so, the monofractal property of the CH magnetic field at scales above 2 Mm
seems to be the most plausible reason why the averaged magnetic flux density,
derived from observations with low resolution, does not correlate with the solar
wind speed: the bulk of energy release dynamics, needed for the solar wind
acceleration, occurs at smaller scales, where the magnetic field structure is
entirely different.

Berger and colleagues (Berger {\it et al.}, 2004) observed a plage area with 0.1
arcsec resolution, and they report that the magnetic flux is structured into
amorphous ribbon-like clusters with embedded knots of enhanced density, which
seems to correspond to the notion of multifractality at small scales.

It is worth to mention that the property of monofractality of solar magnetic
fields was known for a long time (see references in the previous section).
Difficulties of describing magnetic structures with a single fractal dimension
at the entire range of scales was also noticed (see, {\it e.g.}, Tarbell {\it et
al.}, 1990; Janssen {\it et al.}, 2003), which actually is a signature of
multifractality. Fractal analysis of high resolution magnetograms from VTT with
0.4 arcsec spatial resolution for a quiet sun area (Janssen {\it et al.}, 2003)
revealed a break of self-similarity at scales of 1.3 Mm, which is very close to
the scale found in this study.

For an active region and a plage area, our approach for deriving the filling
factor of the magnetic field produced that $f(r) =0.14-0.17$ at $r=0.3$ Mm that
generally agree with earlier reports. According to Tarbell {\it et al.} (1979),
Schrijver (1987), Berger {\it et al.} (1995) and references herein, the magnetic
filling factor is typically inside a range of 10 - 25\%. In active regions,
where the range of solar flares spreads over all observable scales,
multifractality is also present at the same range of scales. This presents
further evidence that energy release dynamics and the multifractality are
mutually related properties of solar magnetic fields.

With the Hinode instrument in operation, many new phenomena related to the CH
dynamics, coronal heating and solar wind acceleration will be discovered now,
when the spatial scales less than 1000 km are available for analysis. Recent
study of the evolution of network magnetic elements (Lamb {\it et
al.}, 2008) proved that processes at sub-resolution scales are of vital
importance for understanding the observed dynamics of magnetic flux. 

Acknowledgements.
Authors thank Spiro Antiochos, Len Fisk, Dennis Haggerty, Marco Velli, Yi-Ming
Wang, Thomas Zurbuchen and the entire LWS/TR\&T Heliospheric Focus Team for
helpful discussions and initiation of this research. We also thank Rob Markel
for valuable assistance in during the inversion process, and anonymous referees
whose criticism and comments led to a significant improvement of the manuscript.
We thank the ACE MAG and SWEPAM instrument teams and the ACE Science Center for
providing the ACE data. SOHO is a project of international cooperation between
ESA and NASA. Hinode is a Japanese mission developed and launched by ISAS/JAXA,
collaborating with NAOJ as a domestic partner, NASA and STFC (UK) as
international partners. Scientific operation of the Hinode mission is conducted
by the Hinode science team organized at ISAS/JAXA. This team mainly consists of
scientists from institutes in the partner countries. Support for the post-launch
operation is provided by JAXA and NAOJ (Japan), STFC (U.K.), NASA (U.S.A.), ESA,
and NSC (Norway). Hinode SOT/SP inversions were conducted at NCAR under the
framework of the Community Spectro-polarimtetric Analysis Center (CSAC;
$http://www.csac.hao.ucar.edu/$). This work was supported by NASA NNX07AT16G
grant, and NSF grant ATM-0716512.

{}

\end{article}


\begin{thebibliography}{}

\bibitem{}
Abramenko, V.I.: 2005, {\it Solar Phys.} {\bf 228}, 29.


\bibitem{}
Abramenko, V.I.: 2008, In: Wang, P.(ed), {\it Solar Physics Research Trends},
 Nova Science Publishers, Inc., New York, 95.


\bibitem{}
Abramenko, V.I., Yurchyshyn, V.B., Wang, H., Spirock, T.J., Goode,  P. R.: 2002,
 {\it Astrophys. J.} {\bf 577}, 487.

\bibitem{}
Abramenko, V.I., Yurchyshyn, V.B., Wang, H., Spirock, T.J., Goode,  P. R.: 2003,
 {\it Astrophys. J.} {\bf 597}, 1135.

\bibitem{}	
Abramenko, V. I., Fisk, L. A., Yurchyshyn, V. B.: 2006, {\it Astrophys.
J.} {\bf 641}, L65.

\bibitem{}
Abramenko, V.I., Yurchyshyn, V.B., Wang, H.: 2008,
 {\it Astrophys. J.} {\bf 681}, 1669.

\bibitem{}
Arge, C.N., Luhmann, J.G., Odstrcil, D., Schrijver, C.J., Li, Y.: 2004, {\it J.
Atmos. Solar Terr. Phys.} {\bf 66}, 1295.

\bibitem{}
Baker, D., van Driel-Gesztelyi, L., Kamio, S., Culhane, J. L., Harra, L.K.,
Sun, J., Young, P.R., Matthews, S.A.: 2008, In: Matthews, S.A.,
Davis, J.M., Harra, L.K. (eds), {\it First Results From Hinode}, {\it ASP Conf.
Ser.} {\bf 397}, 23.

\bibitem{}
Balke, A.C., Schrijver, C.J., Zwaan, C., Tarbell, T.D.: 1993,
{\it Solar Phys.} {\bf 143}, 215.

\bibitem{}
Berger, T.E., Schrijver, C.J., Shine, R.A., Tarbell, T.D., Title, A.M., 
Scharmer, G.: 1995, {\it Astrophys. J.} {\bf 454}, 531.

\bibitem{}
Berger, T. E., Rouppe van der Voort, L. H. M., Lofdahl, M. G., Carlsson, M.,
Fossum, A., Hansteen, V. H., Marthinussen, E., Title, A., Scharmer, G.: 2004,
{\it Astron. Astrophys.} {\bf 428}, 613.

\bibitem{}
Bumba, V., Klvana, M., Sykora, J.: 1995, {\it Astron. Astrophys.} {\bf 298},
923. 

\bibitem{}
Delaboudiniere, J.-P., Artzner, G.E., Brunaud,J., Gabriel, A.H.,Hochedez, J.F.,
Millier, F., {\t ei al.}: 1995, 
{\it Solar Physics} {\bf 162}, 291.

\bibitem{}
Fisk, L.A.: 1996, {\it J. Geophys. Res.} {\bf 101}, 15547.

\bibitem{}
Fisk, L.A.: 2001, {\it J. Geophys. Res.} {\bf 106}, 15849.

\bibitem{}
Fisk, L.A.: 2005, {\it Astrophys. J.} {\bf 626}, 563.

\bibitem{}
Fisk, L.A., Zurbuchen, T. H., Schwadron, N.A.: 1999, {\it Astrophys. J.} {\bf
521}, 868.

\bibitem{} 
Frisch, U.: 1995, {\it Turbulence, The Legacy of A.N. Kolmogorov}, 
Cambridge University Press, Cambridge, 296.

\bibitem{}
Hagenaar, H. J., DeRosa, M. L., Schrijver, C. J.: 2008, {\it Astrophys. J.}
{\bf 678}, 541.

\bibitem{}
Harvey, J., Krieger, A. S., Timothy, A. F., Vaiana, G. S.: 1975, {\it Oss. Mem.
Oss. Arcetri} {\bf 104}, 50.

\bibitem{}
Harvey, J. W., Sheeley, N. R., Jr.: 1979, {\it Space Sci. Rev.} {\bf 23},
139.


\bibitem{}
Harvey, K. L., Harvey, J. W., Sheeley, N. R., Jr.: 1982 {\it Solar Phys.} {\bf
79}, 149.

\bibitem{}
Ichimoto, K., Lites, B., Elmore, D., Suematsu, Y., Tsuneta, S.,
Katsukawa, Y., {\it et al.}: 2008, {\it Solar Phys.} {\bf 249}, 233.


\bibitem{}
Ireland, J., Gallagher, P.T., McAteer, R.T.J.: 2004, In: Dupree, A.K., Benz,
A.O. (eds), {\it Stars as Suns: Activity, Evolution and Planets}, {\it IAU
Symp.} {\bf 219}, 255.


\bibitem{}
Janssen, K., Vogler, A., Kneer, F.: 2003, {\it Astron. Astrophys.} {\bf 409},
1127.

\bibitem{}
Lamb, D. A., DeForest, C. E., Hagenaar, H. J., Parnell, C. E., Welsch, B. T.:
2008, {\it Astrophys. J.} {\bf 674}, 520.

\bibitem{}
Kamio, S., Hara, H., Watanabe, T., Matsuzaki, K., Shibata, K., Culhane, L.,
Warren, H.P.: 2007, {\it Publ. Astron. Soc. Japan} {\bf 59}, S757.

\bibitem{}
Lawrence, J.K., Ruzmaikin, A.A., Cadavid, A.C.: 1993,
{\it Astrophys. J.} {\bf 417}, 805.

\bibitem{}
Longcope, D.W., Parnell, C.E.: 2008, {\it Solar Physics} {\bf 254}, 51. 

\bibitem{}
McAteer, R.T.J., Gallagher, P.T., Ireland, J.: 2005,
{\it Astrophys. J.} {\bf 631}.

\bibitem{}
Meunier, N.: 1999, {\it Astrophys. J.} {\bf 515}, 801.

\bibitem{}
Moreno-Insertis, F., Galsgaard, K., Ugarte-Urra, I.: 2008, {\it Astrophys.
J.} {\bf 673}, L211.

\bibitem{}
Obridko, V. N., Shelting, B. D.: 1989, {\it Solar Phys.} {\bf 124}, 73.

\bibitem{}
Obridko, V., Formichev, V., Kharshiladze, A. F., Zhitnik, I., Slemzin, V.,
Hathaway, D., Wu, S. T.: 2000,
{\it Astron. Astrophys. Trans.} {\bf 18}, 819.

\bibitem{}
Robbins, S., Henney, C. J., Harvey, J. W.: 2006, {\it Solar Phys.} {\bf 233},
265.

\bibitem{}
Scherrer, P.H., Bogart, R.S., Bush, R.I., Hoeksema, J.T., Kosovichev, A.G.,
Schou, J., {\it et al.}: 1995, {\it Solar Phys.} {\bf 162}, 129.

\bibitem{}
Schrijver, C.J.: 1987, {\it Astron. Astrophys.} {\bf 180}, 241.

\bibitem{}
Schrijver, C.J.: 2001, {\it Astrophys. J.} {\bf 547}, 475.

\bibitem{}
Schrijver, C.J., Zwaan, C., Balke, A.C., Tarbell, T.D., Lawrence, J.K.: 1992,
{\it Astron. Astrophys.} {\bf 253}, L1.

\bibitem{}
Schrijver, C.J., Title, A.M.: 2001, {\it Astrophys. J.} {\bf 551}, 1099.

\bibitem{}
Schrijver, C.J., DeRosa, M.L., Title, A.M.: 2002, {\it Astrophys. J.} {\bf
577}, 1006.

\bibitem{}
Schrijver, C.J., DeRosa, M.L.: 2003, {\it Solar Phys.} {\bf 212}, 165.

\bibitem{}
Schroeder, M.: 1991, {\it Fractals, Chaos, Power Laws}, W.H. Freeman and
Company, New York, 429.


\bibitem{}
Schwadron, N.A., McComas, D.J.: 2008, AGU, Fall Meeting, SH13B-1559.

\bibitem{}
Sheeley, N.R., Jr., Harvey, J.W., Feldman, W.C.: 1976, {\it Solar Phys.}
{\bf 49}, 271.


\bibitem{}
Shimojo, M.: 2008, AGU, Fall Meeting, SH22A-0836.


\bibitem{}
Suematsu, Y.: 2008, AGU, Fall Meeting, SH41B-1623.

\bibitem{}
Tarbell, T.D., Title, A.M., Schoolman, S.A.: 1979, {\it Astrophys. J.} {\bf
229}, 387.

\bibitem{}
Tarbell, T., Ferguson, S., Frank, Z., Shine, R., Title, A., Topka, K., Scharmer,
G.: 1990, In: Stenflo, J.O. (ed.), {\it Solar Photosphere: Structure,
Convection, and Magnetic fields}, {\it IAU Symp.} {\bf 138}, 147.

\bibitem{}
Tsuneta, S., Ichimoto, K., Katsukawa, Y., Nagata, S., Otsubo, M.,
Shimizu, T., {\it et al.}: 2008, {\it Solar Phys.} {\bf 249}, 167.

\bibitem{}
Vr\u snak, B., Temmer, M., Veronig, A.M.: 2007, {\it Solar Phys.}
{\bf 240}, 315.

\bibitem{}
Wang, Y.M., Sheeley, N.R.: 1990, {\it Astrophys. J.} {\bf 355}, 726.

\bibitem{}
Wang, Y.M., Sheeley, N.R.: 1991, {\it Astrophys. J.} {\bf 375}, 761.

\bibitem{}
Wang, Y.M., Lean, J., Sheeley, N.R.: 2000, {\it Geophys. Res. Lett.}
{\bf 27}, 505.

\bibitem{}	
Wiegelmann, T., Solanki, S.K.: 2004, {\it Solar Phys.} {\bf 225},
227.

\bibitem{}	
Wiegelmann, T., Xia, L. D., Marsch, E.: 2005, {\it Astron. Astrophys.}
{\bf 432}, L1.


\end{thebibliography}
\end{document}